# Dynamics in poly(propylene glycol) and silica oxide nanoparticles composite glassforming system

M. Głuszek,[1] A. Antosik,[1] R. Żurowski,[1] M. Szafran,[1] S.J. Rzoska,[2] M. Zalewski,[2] E. Pawlikowska[2] and S. Starzonek[2a)]

[1]*Department of Chemical Technology, Warsaw University of Technology, ul. Noakowskiego 3, 00-664 Warsaw, Poland*

[2]*Institute of High Pressure Physics, Polish Academy of Sciences, ul. Sokołowska 29/37, 01-142 Warsaw, Poland*

Results of broadband dielectric spectroscopy and rheological studies of poly(propylene glycol) + $SiO_2$ nanocomposites are presented. They show that the dynamics in high-concentrated composite is determined by confinement and adsorption effects, resulting from interactions of confined polymers correspond with the host system at the interface between PPG and solid nanoparticles. The evolution of relaxation times follows the clear Vogel-Fulcher-Tammann pattern, what is proved by the supplementary activation energy temperature index analysis. The strong influence of nanoparticles on the transitional dynamics and the fragility in the ultraviscous domain was noted.

**I. INTRODUCTION**

Glass transition and related phenomena are considered as one of greatest challenges of the condensed matter physics, soft matter physics and material engineering [1]. One of its most fascinating features are far previtreous effects for dynamic properties, starting even ca. 200 K above the glass temperature ($T_g$). They exhibit a set of universal patterns, sharing between glass forming systems, notably different at the microscopic level. One can recall here [1-5]: (i) the non-Debye distribution of relaxation times, (ii) the non-Arrhenius evolution of relaxation times or viscosity, (iii) the 'magic' dynamic ergodic – nonergodic crossover most often at $\tau(T_B) \approx 0.1 \mu s$ for relaxation time and $\eta(T_B) \approx 1 kPoise$ for viscosity, (iv) associated with the latter decoupling between translational and orientational properties, (v) below the dynamic crossover the secondary *beta* relaxation process emerges. One of paths for approaching the key point of this mystery can be studies in 'innovative glass formers', revealing some usually hidden features of the phenomenon.

This report is related to such studies focusing on ultraviscous poly(propylene glycol) (PPG) and silica oxide ($SiO_2$) nanoparticles (NPs) based nanocomposite systems. The glassy dynamics

[a)] Corresponding author. Institute of High Pressure Physics, Polish Academy of Sciences, ul. Sokołowska 29/37, 01-142 Warsaw, Poland. E-mail address: starzoneks@unipress.waw.pl (S. Starzonek)

of poly(propylene glycols) (PPGs) has been already broadly studied [5-10], yielding from this compound a good candidate as the host for nanocomposite system. Regarding composites, the evidence for glassy dynamics is still limited but it is clear that they seem to constitute a separate class of glass formers created by beneficial combination of the ultraviscous glass former and solid particles [10-12]. Results presented below are based on broadband dielectric spectroscopy (BDS) and supplementary rheological tests, two methods essential for getting insight into dynamics of glass formers [4, 5].

## II. EXPERIMENTAL

### A. BROADBAND DIELECTRIC SPECTROSCOPY AND PREVITREOUS DYNAMICS

The broadband dielectric spectroscopy (BDS) offers a simultaneous insight into dynamics for even 15 decades in frequency/time thus yielding a unique tool for studying previtreous phenomena in glass formers [5]. Moreover, BDS can be relatively easy implemented for high pressure studies although in this case the terminal frequency is limited to now more than 10 MHz, so far [11]. Experimental data from BDS scans can be shown in few representations focusing on translational and orientational processes. The latter is associated with the complex dielectric permittivity as a function of a frequency $\varepsilon^*(f) = \varepsilon'(f) - i\varepsilon''(f)$. The imaginary part of permittivity gives direct access to the distribution and values of the primary relaxation time, namely $\tau = 1/2\pi f_{peak}$ where $f_{peak}$ is the frequency coordinate of the primary dielectric loss curve. The distribution of relaxation times is associated with slopes of loss curves and alternatively can be estimated also from $\varepsilon''(f)$ fittings via a set of formulas, amongst which Havriliak – Negami one is the most general [5]. The temperature evolution of dynamic properties, amongst which the primary relaxation times plays the key role in the ultraviscous domain of glass formers and expressed by the general super-Arrhenius (SA) equation [2, 5]:

$$\tau(T) = \tau_0 \exp\left(\frac{\Delta E_a(T)}{RT}\right) \qquad (1)$$

where $\Delta E_a(T)$ denotes the appartent activation energy and $\tau_0$ is the prefactor which can range between $10^{-11}$ s and $10^{-16}$ s in different systems.

In the simplest 'classical' case the apaprent acitivation energy is constant, i.e. $\Delta E_a(T) = \Delta E_a = const$, and only in this case the activation energy can be easily determined as $\Delta E_a = R \ln \tau(T) / d(1/T)$. For the temperature – dependent apparent activation energy $R \ln \tau(T) / d(1/T) = \Delta H_a(T)$, where $\Delta H_a(T)$ is the apparent activation enthalpy. The routine for non-biased determing of $\Delta E_a(T)$ has been developed in Refs. [2, 3, 13]. The application of the SA eq. (1) for portraying $\tau(T)$ experimental data is still not possible, and 'ersatz relations' are used. The most important and popular is the Vogel-Fulcher-Tammann (VFT) dependence [1, 4, 5]:

$$\tau(T) = \tau_0 \exp\left(\frac{D_T T_0}{T - T_0}\right), \quad (2)$$

where $D_T$ is the fragility strength coefficient related to the temperature of the path of the approaching the glass transition, $T_0$ is the extrapolated VFT singular temperature, located in the solid glass phase, below the glass temperature $T_g$.

In the analysis experimental data are most often presented via the Arrhenius plot $\log \tau(T)$ or $\ln \tau(T)$ vs. $1/T$ and the nonlinear fit via eq. (2) is subsequently directly implemented [1, 4, 5]. However, such procedure introduces notable biased factors which can yield 'effective' and non-physical values of parameters. This can be associated with the erroneously selected fitting domain and the fact that the given relations can appear non-optimal or non-physical. Next issue is associated with the nonlinear fitting procedure, which also may yield non-physical and effective values of parameters. In refs. [14, 15] the simple recipe for solving these problems was proposed. Taking derivatives of eqs. (1) and (2) and merging resulted equations one can obtain the following dependence [14, 15]:

$$\left[\frac{d \ln(\tau)}{d (1/T)}\right]^{-0.5} = (\Delta H'_a)^{-0.5} = A - \frac{B}{T} \quad (3)$$

Thus the appearance of the linear domain at the plot defined by eq. (3) indicates the domain of a possible validity of the VFT parameterization and the subsequent basic linear regression fit yields optimal values of key parameters in eq. (2): $T_0 = |B/A|$ and $D_T = 1/|AB|$. In fact, the plot $d \log \tau(T)/d(1/T)$ vs. $(1/T)$ was earlier used by Stickel et al. [13], but focusing solely on the estimation of the location of the so called dynamic crossover temperature and no link to the activation enthalpy was indicated [5, 16].

In recent decades few other equations offering a reliable description of the temperature evolution of relaxation time appeard and they seems to offer equally reliable description as VFT, including the support of the preliminary derivative-based analysis. Hence the question remains if the VFT equation is indeed the optimal choice for the given system. This basic problem can be solved the supplementary analysis using the activation energy temperature index $I_{DO}(T)$ [2, 3, 13]

$$I_{DO}(T) = -\frac{d \ln E_a(T)}{d \ln T}. \qquad (4)$$

The form of this index for different relations describing dynamics in the ultraviscous/ultraslowing domain is notable different, thus enabling the ultimate and unequivocal selection. In the case of the VFT dependence one obtains [2]:

$$I_{DO}^{VFT}(T) = \frac{T_0}{T-T_0}. \qquad (5)$$

The determination of $I_{DO}(T)$ values from $\tau(T)$ experimental data requires the knowledge of the apparent activation time $E_a(T)$ in prior. The non-biased way of calculating $E_a(T)$, based on the numerical solution of a differential equation, has been recently proposed in refs. [2, 3, 13]. Subsequently, it was also discovered that the form of the evolution of the index may be 'universal feature' of glass forming dynamics $I_{DO}^{-1}(T) = a + bT$ [2], i.e. shows the linear behavior for arbitrary glass formers. This feature was shown for the set of ca. 50 glassy systems, ranging from low molecular weight liquids to polymers, liquid crystals and plastic crystals [2, 3]. In the case of the VFT equation one obtains:

$$[I_{DO}^{VFT}(T)]^{-1} = \left(\frac{1}{T_0}\right)T - 1.$$

In ultraviscous/ultraslowing regime the evolution of translational and orientational motions becomes decoupled, what can be detected via the Debye–Stokes– Einstein relation due to its shift to the fractional form (FDSE) [11]:

$$\sigma_{DC}(T)[\tau_a(T)]^S \qquad (6)$$

where $\sigma_{DC}$ and $\tau_\alpha$ stands for DC-conductivity and the primary relaxation time.

The experimental evidence indicates that generally FDSE or FSE behavior with non-zero fractional exponents takes place in the ultraviscous/ultraslow dynamical domain for $T_B < T(\tau) < T_g$. For $T > T_B$ the crossover to Debye–Stokes– Einstein (DSE) occurs when $S = 1$. Study of FDSE behavior is one of the key tools for obtaining insight into dynamic heterogeneities, called cooperatively rearranged regions near $T_g$ [4, 5, 11]. They are recognized as the most probable reason for the universal patterns in the ultraviscous domain. It was shown that $S = \Delta H_a^\sigma(T)/\Delta H_a^\alpha(T) = m_P^\sigma/m_P^\alpha$, where $m$ is the fragility index, $P = const$. The most precise calculation in Ref. [11] can be found.

Basing on existing reference experimental results [5, 8, 9] for ultraviscous PPGs one can note few types of relaxation processes in this system: the normal mode (NM) related to the polymer chains interactions (because PPG is a type-A polymer), $\alpha$-process/segmental mode (SM) defined as actions of the polymer segments, $\beta-$ and $\gamma-$ processes associated with deep glass motions. So far, the temperature evolution of the $\alpha$-relaxation is described by the single VFT eq. (3).

For the given report measurements of the complex dielectric permittivity $\varepsilon^*(f)$ were performed using the broadband dielectric spectrometer (BDS) Novocontrol Alpha Analyzer over a frequency range from $10^{-2}$ to $10^7$ Hz . This enabled 6 digits resolution in dielectric measurements. The temperature was controlled by the Quattro System (Novocontrol) with stability better than 0.2 K [5] . Samples were placed in the measurement capacitor made from Invar, with a gap $d = 0.2$ mm and a diameter $2r = 20$ mm. The Teflon ring was used as the spacer. The latter and the macroscopic gap of the capacitor made it possible to avoid bubbles, distorting results.

**B. RHEOLOGY**

Rheological measurements are very useful technique for studying viscosity as a function of the applied shear rate or shear time. The fluids behavior are divided into Newtonian and non-Newtonian flow due to the rheological characteristics. Newtonian with stable rheology independents of time and the non-Newtonian fluids with rheological characteristics dependent on shear rate or time constitute an essential types of fluids. The basic law of viscosity describing the Newtonian flow behavior of an ideal liquid is as follow [17, 18]:

$$\tau = \eta_0 \dot{\gamma}, \qquad (7)$$

where $\eta_0$ is a Newtonian viscosity, $\dot{\gamma}$ is the shear rate and $\tau$ is the shear stress. The velocity gradient $[\mathrm{d}u/\mathrm{d}y]$, which is termed the shear rate $\dot{\gamma}$ is given by [17, 18]:

$$\dot{\gamma} = \frac{\mathrm{d}u}{\mathrm{d}y} = \frac{\mathrm{d}}{\mathrm{d}y}\left[\frac{\mathrm{d}l}{\mathrm{d}t}\right] = \frac{\mathrm{d}}{\mathrm{d}t}\left[\frac{\mathrm{d}l}{\mathrm{d}y}\right]. \qquad (8)$$

The term $[\mathrm{d}l/\mathrm{d}y]$, represents the deformation of the material and is defined as the shear strain $\gamma$. The shear rate is in equilibrium with the value of the gradient of the rate of particulate or grains on a fixed point of the fluid [17, 18]. In classifying the description of non-Newtonian fluids [17], there are pseudoplastic, dilatants [19], Bingham plastics, pseudo-plastic with a yield stress [20], thixotropic, rheopectic [21, 22] and viscoelastic fluids [23, 24]. There are a number of factors which affect dopant polymer interactions which influence on rheological properties of suspensions. The length and molecular structure of the chain are the main factors determining the rheology because of the possibility to molecules linking together and consequently occupying a large part of space compared to its atomic dimensions [17, 25]. The viscosity of a polymeric system decreases with increasing temperature due to the greater free volume available for molecular motion at the higher temperature [26, 27]. It is also well known that physical properties change in parallel to solid loading [28]. Highly concentrated materials show higher viscosity because interaction between polymeric chain and particles are stronger (attractive van der Waals forces, repulsive electrostatic forces, Brownian motion, force-distance relation). In case of $SiO_2$/PPG suspensions hydroxyl group in carrier fluid could merge with the silanol group on the surface of silica oxide to form hydrogen bond, so links among particles are created. When shear rate reaches a critical value, particles start to move from the initial position and the layer structure is damaged. As particles move, they begin to aggregate and form the particle clusters along the direction of shear due to the links among particles. With the increase of shear rate, the particle clusters rise and the viscosity becomes great [28]. The dependencies of rheological properties strongly correspond also to particle size of solid particles [29]. The higher particle size leads the smaller free space and distance between components. This resulting in stronger internal friction and viscosity values. For examination samples rheological properties the rotational rheometer KinexusPro (Malvern) with plate-plate system was used. The diameter of the top rotating plate was 20 mm, and of the bottom stationary plate was 100 mm. The gap was set to $d = 0.1$ mm for pure PPG and $d = 0.7$ mm for suspensions. The viscosity measurements were taken at temperatures $T_1 = 273$ K and $T_2 = 301$ K in range from $\dot{\gamma} = 0.1$ s$^{-1}$ to $\dot{\gamma} = 3000$ s$^{-1}$.

**C. TESTED COMPOSITE SAMPLE**

Regarding preparations of suspensions amorphous spherical silica oxide nanopowder with a diameter $2r = 100$ *nm* as a solid phase and poly(propylene glycol) H[OCH(CH3)CH2]$_n$OH, where n = 35 with an average molecular weight $M = 1000$ g·mol$^{-1}$ (PPG1000) as a carrier liquid were used.

TABLE I. Basic characterization of tested composite samples

| Sample | Composition KE-P10 vs. PPG | | Dynamic type |
|---|---|---|---|
| | Volume fraction (%v/v) | Mass fraction (%wt/wt) | |
| $S_0$ – PPG1000 | 0/100 | | Newtonian |
| $S_1$ – Composite | 50/50 | 66.1/33.9 | Non-Newtonian |

A specific surface area of SiO$_2$ examined by a Brunauer-Emmett-Teller adsorption isotherm, ASAP 2020 (Micromeritics) was $S_{BET} = 132.2$ m$^2$·g$^{-1}$. A density equals $\rho = 1.96$ g·cm$^{-3}$ was measured using helium pycnometer AccuPyc 1340 Pycnometer (Micromeritics).

The silica oxide nanopowder was supplied by Nippon Shokubai and the PPG1000 by Sigma-Aldrich. TAB. 1. presents the liquid nanocomposites and their abbreviations. The suspension was obtained by mixing dry $SiO_2$ nanopowder with PPG using mechanical stirrer by 1 hour. No sedimentation for at least 24 hours was observed, hence the final colloid did not contained any additional stabilizing agent.

FIG. 1. shows dependences of viscosity versus the shear rate coefficient for pure PPG1000 ($S_0$) and studied composite ($S_1$). In case of PPG1000 the viscosity was independent of the shear rate and it equaled c.a. 0.14 Pa·s. Such dependency is typical for Newtonian fluids which indicates that molecules interact within the thermodynamic equilibrium state. On further increasing the amount of $SiO_2$ in the system the average distance between nanoparticles decreases strengthening internal interactions. The distance between nanoparticles can be estimated by the relation [30]:

$$h = D\left[\left(\frac{1}{3\pi\phi} + \frac{5}{6}\right)^{0.5} - 1\right], \qquad (6)$$

where $h$ is a distance between ceramic nanoparticles, $D$ is the particles diameter and $\phi$ is a solid loading. Its implementations yields 2.25 nm (for suspension $S_1$). It is notable that for this nanocomposite the behavior is clearly non-Newtonian for any tested shear rates. The rapid increase of viscosity from 18 Pa·s to 978 Pa·s can be linked to the shear flow induced arrangement of the ceramic nano-powder within the host, leading to the increase internal friction forces and jamming the fluid flow at shear rate in range from 2.63 $s^{-1}$ to 34.7 $s^{-1}$. The subsequent shear thickening phenomenon can be interpreted as the result of the order-disorder transition [30], hydrodynamic clustering [31], partial flocculation [32] and dilatancy [33, 34].

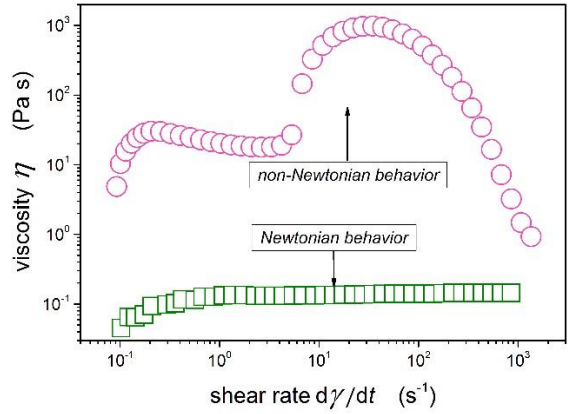

FIG. 1. Basic results showing the type of the rheological behavior for tested ultraviscous composites. The horizontal domains indicate Newtonian domains, i.e. shear rate independent flow.

### III. RESULTS AND DISCUSSION

The main part of Fig. 2. shows the evolution of relaxation time for types $S_0$ (Newtonian) and $S_1$ (non-Newtonian) systems. The plot contains results both from rheological and BDS studies. The segmental mode ($\alpha$-process) in both cases exhibit nonlinearity what indicates-the non-Arrhenius behavior. For this relaxation mechanism there are no detectable differences between pure PPG and its silica oxide composites. The normal mode (NM) relaxation exists in pure poly(propylene glycol) with the molecular mass higher than $M = 1000$ g/mol, however its hallmark are relatively very weak [5, 35-37]. The distorting impact of the electric conductivity is visible. It is notable that the relaxation processes for pure PPG1000 and $S_1$ composite coincide in the high temperature region which proves that the α-process and the normal mode exist despite the confined space.

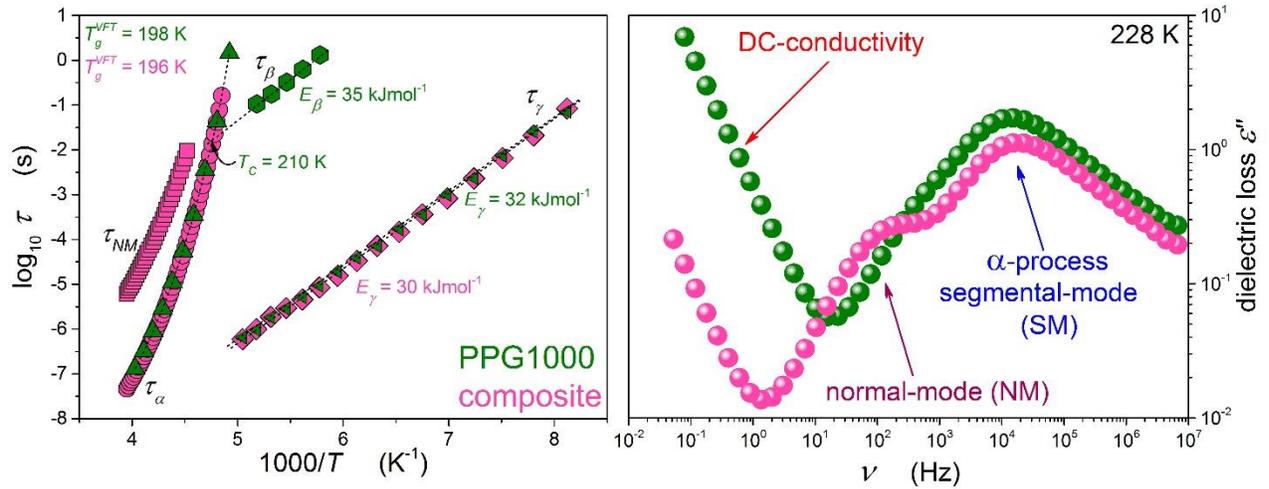

FIG. 2. The temperature evolution of key relaxation times in the non-Newtonian composite ($S_1$) and pure PPG1000 ($S_0$) host (Newtonian) system obtained from the broadband dielectric spectroscopy. The right part presents the dielectric loss spectra with relaxation processes: low-frequency normal mode (NM) and high-frequency segmental mode (SM) for both samples.

Both modes relaxations: normal (NM) and segmental (SM) can be reliable parameterized via the VFT behavior when the direct fit via eq. (2) is implemented. In both cases system become 'stronger' on approaching the glass transition but this change is more pronounced for the composite system. The strong shift is notable, towards higher temperatures for the latter case.

The nature of NM relaxation can be explained basing on possible hydrogen bounds between nanoparticles surface and polymer chains, but another mechanisms related to end-to-end polymer chain motions for PPGs ($M$ = 2000 g/mol) were also indicated [37]. Schönhals and co-authors based on matched light scattering and BDS experiments showed that PPGs confined within amorphous $SiO_2$ nanopores exhibit (slow) low-frequency relaxation process called normal mode (NM) [8, 9]. They studied bulk PPGs and the influence of different nanopores diameters, receiving overlapping the results. Roland and co-workers reported results of similar studies for PPG1000/$SiO_2$ nanoparticles mixture, exhibiting the Newtonian behavior. In their report the normal mode relaxation is recalled as the global one and the segmental mode as the local one [10]. The segmental mode dynamics is stimulated by these same reasons what the pure PPG for the high concentration of silica oxide nanoparticles. It is notable that an asymmetric broadening of dielectric spectrum did not occur for observed effects what is proved by similar positions of maxima of loss curves peaks.

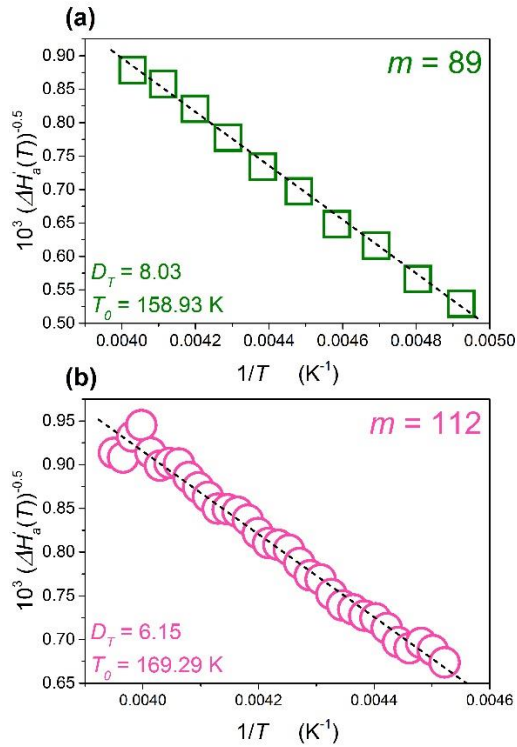

FIG. 3. The linearized plots of the temperature dependences of the reduced apparent activation enthalpy $\Delta H'_a(T)$ for (a) pure poly(propylene glycol) $M = 1000$ g·mol$^{-1}$, (b) its silica oxide nanocomposite (50:50), focused on the validity of the VFT relation (3). The fragility index $m$, calculated by Böhmer equation, equals (a) 89 and (b) 112, respectively.

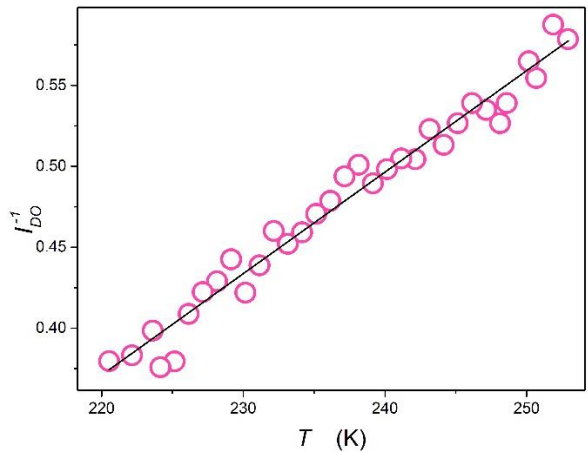

FIG. 4. The evolution of the activation energy temperature index for PPG 1000 based nanocomposite, determined using the protocol proposed in refs. [2,13]. It is described via $I_{DO}^{-1} = a + bT$, where $a = -1.01$ and $b = 0.0063$, showing the clear agreement with the form expected for the VFT equation. One can estimate the VFT singular temperature

However, results of the given report shows that the primary relaxation dynamics slows down in the S$_1$ sample, in comparison with the host PPG sample what is visible also at the impact on the glass transition temperature value. One can conclude that the molecular dynamics in PPG composite is regulated by surface and confinement effect what come from the non-monotonic dependence of the relaxation times. This results give a clear evidence that the length scale is significant for the dynamics of glass forming systems. The latter mechanism seems to be possible when taking into consideration the size of nanoparticles (c.a. 100 nm) and the average polymer chain length (c.a. 25 nm). Moreover, intermolecular interactions are produced by oxides on nanoparticles surface. The distance between silica oxide nanoparticles calculated from the Eq. (8) corresponds with the nanoporous experiments which were performed in Ref. [8, 9]. All these can lead to the *H*-bounds interactions. Notable is also the change in such basic characteristics as the fragility defined as $m = \left[d\log_{10}(\tau)/d(T_g/T)\right]_{T=T_g}$ which can be deduced from results presented in Fig. 3 via the relation $D_T = 590/(m-16)$ [3, 5, 38], what yield $m = 89$ for S$_0$ and $m = 112$ for S$_1$. This means, that presented studies revealed the clear evidence of two dynamic domains, strongly influenced by the presence of SiO$_2$ nanoparticles.

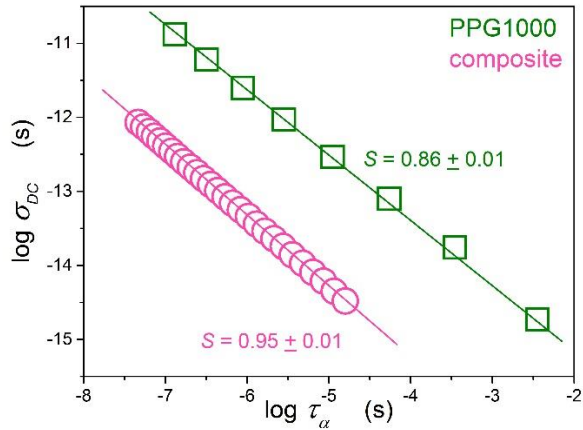

FIG. 5. Fractional Debye-Stokes-Einstein (FDSE) test for studied PPG1000 (squares) and composite (rounds). The fraction *S* describing the slope of the data (straight lines) equals 0.86 and 0.95 for S$_0$ and S$_1$, respectively.

Fig.5. presents the interplay between transitional ($\sigma_{DC}(T)$) and orientational ($\tau_\alpha(T)$) dynamics described by the fractional Debye-Stokes-Einstein relation (FDSE) given by Eq. (6) for $T_B < T < T_g$

regime. The exponent for the pure PPG $S \approx 0.86$ and for the composite $S \approx 0.95$. The straight line denotes the negligible decoupling between DC-conductivity and relaxation time take place. In studied systems the main role by transitional dynamics is played. This behavior is correlated to the higher concentration of ions and interplay between nanoparticles surface and polymer chains for composite system .

**IV. CONCLUSION**

In summary, results above show that doping nanoparticles into polymer systems strongly influence dynamics and viscoelastic properties of PPG based glass former. Dielectric study revealed the significant impact of $SiO_2$ nanoparticles on the creation of global interactions within the composite system. Their origins can results from hydrogen bounding between nanoparticles surface and polymer chains. The inter-chain interactions seem to play a scant role for PPGs with molecular mass less than $M = 2000$ g/mol. In previous reports [8, 9, 35], the correlation between confined and high-concentrated system takes place. Indeed, in studied case ($S_1$) the polymer (PPG) chains were confined between silica oxide nanoparticles, what let create an intermolecular interactions. On the other hand, NPs exhibit no impact on the segmental (local) relaxation. Our report proves that the molecular dynamics in high-concentrated composite is determined by confinement and adsorption effects. This and earlier results from interactions of confined polymers correspond with the host system at the interface between both. The most notable is the very strong influence on parameters describing the transitional-orientational decoupling (FDSE) and fragilities in the ultraviscous/ultraslow domain.

**ACKNOWLEDGEMENTS**

The research carried out by was supported via the National Centre for Science (NCN, Poland) grant, ref. 2016/21/B/ST3/02203 (SJR, ADR, SS). Other co-authors were supported from the budgetary resources for science for years 2015-2019 (by Ministry of Science and Higher Education) within Diamond Grant program (Agreement No. 0143/DIA/2015/44). MG, RŻ are grateful the Warsaw KNOW Consortium for the scholarship.